# Determination of spin-dependent Seebeck coefficients of CoFeB/MgO/CoFeB magnetic tunnel junction nanopillars

*N. Liebing[1], S. Serrano-Guisan[1*], K. Rott[2], G. Reiss[2], J. Langer[3], B. Ocker[3] and H. W. Schumacher[1]*

1) Physikalisch-Technische Bundesanstalt, Bundesallee 100, D-38116 Braunschweig, Germany.
2) University of Bielefeld, Department of Physics, Universitätsstr. 25, 33615 Bielefeld, Germany.
3) Singulus AG, Hanauer Landstrasse 103, D-63796 Kahl am Main, Germany

* Corresponding author:

E-mail: santiago.serrano-guisan@ptb.de

Phone: +49 (0)531 592 2439, fax: +49 (0)531 592 69 2439

## Abstract

We investigate the spin-dependent Seebeck coefficient and the tunneling magneto thermopower of CoFeB/MgO/CoFeB magnetic tunnel junctions (MTJ) in the presence of thermal gradients across the MTJ. Thermal gradients are generated by an electric heater on top of the nanopillars. The thermo power voltage $V_{TP}$ across the MTJ is found to scale linearly with the heating power and reveals similar field dependence as the tunnel magnetoresistance. The amplitude of the thermal gradient is derived from calibration measurements in combination with finite element simulations of the heat flux. Based on this, large spin-dependent Seebeck coefficients of the order of $(240 \pm 110)$ µV/K are derived. From additional measurements on MTJs after dielectric breakdown, a tunneling magneto thermopower up to 90% can be derived for 1.5 nm MgO based MTJ nanopillars.

**Article**

Combination of *spintronics*[1] and thermoelectricity in magnetic nanostructures[2,3] can be a promising approach to develop future *pure spin-based* devices with applications in sensing and magnetic data storage. The recent discovery of the Spin Seebeck Effect[4,5,6] suggesting the possibility to generate pure spin currents by means of thermal gradients[7] boosted this new emerging field of *spin caloritronics*[8]. However, a deep understanding of thermoelectric voltage signals in nanoscale magnetic structures is still required. Magnetic tunnel junctions (MTJs) with large tunnel magneto resistance (TMR) ratios have become very important devices for *spintronics*[9] and are also suitable to investigate to investigate *spin caloritronics*. Recent experiments have demonstrated the generation of thermally induced spin dependent voltage signals in CoFeB/MgO/CoFeB MTJs[10,11] and in ferromagnet/insulator/semiconductor tunnel structures[12] yielding large spin dependent signal contributions up to 400 μV[12].

Here, we measure and compare the tunneling magneto thermopower (TMTP) of a number of nominally identical CoFeB/MgO/CoFeB nanopillars under the presence of a thermal gradient $\Delta T_{MTJ}$ across the MTJs. We derive $\Delta T_{MTJ}$ by combining calibration measurements with finite element modeling of the heat flux in the nanopatterned devices. Based on this, the spin-dependent Seebeck coefficients of the MTJs and their uncertainty are derived and compared to theoretical predictions.

The experiments are carried out on MTJ stacks consisting of 3nm Ta, 90nm Cu, 5nm Ta, 20nm PtMn, 2nm $Co_{60}Fe_{20}B_{20}$, 0.75nm Ru, 2nm $Co_{60}Fe_{20}B_{20}$, 1.5 nm MgO, 3nm $Co_{60}Fe_{20}B_{20}$, 10nm Ta, 30nm Cu, and 8nm Ru.[13] Details on the sample structure and magnetic properties can be found elsewhere.[14,11] The stack is patterned into a Cu bottom contact (BC) with 160 nm × 320 nm wide elliptic MTJ nanopillars on top. Fig. 1 (a & b) shows a cross sectional sketch of the layer structure and the contact layout in a Scanning Electron Microscope (SEM) picture. Top contacts (TC) to the MTJ nanopillars allow both measurements of the TMR and the thermoelectric voltage $V_{TP}$. Thermal gradients across the pillars are generated by current application through a heater line (HL) situated on a 160 nm thick $Ta_2O_5$ dielectric. For thermoelectric measurements DC heater currents up to $I_{heat}$ = 60 mA are applied through the HL while $V_{TP}$ between BC and TC of the MTJs is measured. Here, a positive $V_{TP}$ corresponds to a positive voltage at the TC. Note that in the experiments the additional magnetic field generated by application of $I_{heat}$ is always compensated by an external bias field $H_{bias}$.

A quantitative analysis of thermoelectrical signals requires the knowledge of thermal gradients in the nanopillars. Therefore, first the temperatures of HL and BC are determined by calibration measurements using the resistance $R$ of HL and BC as a probe. The temperature dependence of $R_{HL}$ is determined by heating the chip on a variable temperature probe station in a temperatures range of 298 - 333 K. Fig. 1(c) shows the temperature dependence of $R_{HL}$. Linear regression of $R_{HL}$ allows to determine the temperature coefficient of the HL as $\alpha$ = (2.5±0.3) ·$10^{-3}$ $K^{-1}$, in good agreement with reported values of $\alpha$ for Au wires of similar dimensions.[15]

Using this calibrated HL thermometer the temperature of the HL during heating is measured. Fig. 1 (d) shows the resistance change $\Delta R$ of HL and BC under application of a heating power $P_{heat}$ applied to the HL. $R_{HL}$ increases linearly with $P_{heat}$. From $R_{HL}$ the HL temperature increase $\Delta T_{HL}(P_{heat})$ is derived by $R_{HL}$ = $R_{HL,0}$ (1+$\alpha$·$\Delta T_{HL}$), where $R_{HL,0}$ is the room temperature resistance. Using the above $\alpha$ we find a maximum $\Delta T_{HL}$ ≈ 21 K for maximum $P_{heat}$ = 60 mW for the given device. Note that, in parallel, no change of resistance of the BC line was observed although a similar $\alpha$ can be expected (full dots in Fig. 1d). From this we can assume that the Cu BC line acts as an efficient heat sink and the BC temperature does not significantly increase during our experiments. Based on this the temperature distribution within the nanopillar is computed using a commercial finite element solver.[16]

The heat flux in the devices is simulated using a two dimensional cross section model of the nanopillar structure including all contacts and insulating layers. The position of the simulated cross section through the sample is marked in Fig. 1 (a) by the red line. The 2D cross section contains all critical parts of the devices such as HL, TC, MTJ nanopillar, BC, dielectric layers, and substrate. A nanopillar width of 320 nm was used in the simulations. A simulated temperature map of the center part of this 2D cross-section is shown in Fig. 1 (e). For the simulation the temperature at the bottom of the BC was set to 300 K (room temperature) to take into account the negligible temperature increase within the BC. To consider the efficient thermal flux out of the device via the gold TC also the outer boundary of the TC on the right hand side was set to 300 K. The thermal properties of all components of the MTJ stack are based on literature values as given in Table I. [15,17,18,19,20,21,22,23] Whenever available, thin film values of relevant parameters were used. Otherwise bulk values were considered as marked in the table.

The simulation of Fig. 1(e) uses $P_{heat}$ = 60 mW corresponding to a HL temperature of 321 K. The simulations show that the dominant temperature drop occurs across the 160 nm dielectric between HL and TC. This results in a significantly smaller temperature drop between TC and BC of less than 1 K (cp. color scale bar). This blocking of the heat flux by the dielectric can also be observed in the vicinity of the MTJ nanopillar. Here the temperature of the TC directly above the MTJ nanopillar is reduced compared to the surrounding parts of the TC showing a more efficient heat flux through the MTJ than through the surrounding dielectric. The relevant temperature drop for the interpretation of spin-dependent Seebeck measurements is the temperature difference across the MgO barrier $\Delta T_{MTJ}$ between the CoFeB pinned layer (bottom) and free layer (top). $\Delta T_{MTJ}$ is derived from the simulated temperature profile in the center of the MTJ nanopillar in growth direction. The corresponding temperature profile is shown in Fig. 2(f). For maximum heating power the simulations yield $\Delta T_{MTJ} \approx$ 38 mK. Simulations for different $P_{heat}$ show the expected linear scaling of $\Delta T_{MTJ}$.

Concerning the uncertainty of the above $\Delta T_{MTJ}$ different contributions have to be taken into account. From the fitting curves of the temperature calibrations we can estimate an uncertainty of the temperature input parameters of the simulations of about 50 mK/mW resulting in an uncertainty of $\Delta T_{MTJ}$ of about 0.1 mK/mW. Furthermore errors in the thickness of the MTJ layer stacks and in the device geometry should result in an additional uncertainty of a few percent %. The dominant contribution of the uncertainty budged can, however, be attributed to the input material parameters listed in Table I. The use of bulk values as well as the uncertainty of the known thin film values could, e.g., easily sum up to a total uncertainty of 40%. This large error must be taken into account for the following determination of the spin dependent Seebeck coefficient based on magneto thermoelectric measurements.

For this purpose, 8 nominally similar MTJ nanopillars were studied by magneto transport and thermoelectric measurements. For each sample, the easy axis of both the tunnel magneto resistance (TMR) and the tunnelling magneto thermopower (TMTP) were measured. Results are summarized in Table II. Fig. 2 shows the easy axis TMR loop (a, c) and the TMTP loops (b, d) for two different MTJ nanopillars (MTJ-1 (a,b) and MTJ-3 (c,d)). The nanopillars typically show TMR ratios between 70 and 140 %, and resistance area products of the order of ~ 17 $\Omega\mu m^2$. Fig 2(b, d) shows the measured $V_{TP}$ for four different values of $P_{heat}$. $V_{TP}$ shows the same field dependence as the TMR and the devices thus reveal a clear spin dependent TMTP. For all devices the TMTP loops show the same magnetic field dependence as the TMR pointing out an intimate relationship of TMTP and the relative magnetization orientation of the magnetic layers. Furthermore, in all TMTP loops a lower thermopower voltage $V_{TP}(P)$ is measured in the parallel (P) state than in the antiparallel (AP) state $V_{TP}(AP)$. This behavior is independent of $P_{heat}$ as shown in Fig. 2(e). Here $V_{TP}$ (AP,P) is plotted as function of $P_{heat}$. Both $V_{TP}(P)$ and $V_{TP}(AP)$ scale linearly with $P_{heat}$ and hence with $\Delta T_{MTJ}$ as expected. As a result also the spin dependent change of the thermopower $\Delta V_{TP} = V_{TP}(AP) - V_{TP}(P)$ scales linearly with $P_{heat}$ and thus with $\Delta T_{MTJ}$.

- 5 -

Based on this the spin-dependent Seebeck coefficient $\Delta S_{MTJ}$ of the MgO MTJ nanopillars is derived using $\Delta S_{MTJ} \equiv S_{MTJ}(AP) - S_{MTJ}(P) = \Delta V_{TP} / \Delta T_{MTJ}$. The resulting values are listed in Table II. By averaging the values of the nominally identical devices we obtain an average value of $<\Delta S_{MTJ}> = (240 \pm 110)$ µV/K. Note that this error reflects the *statistical* uncertainty from averaging. The value is in agreement with recent *ab-initio* studies for this material system which also predicted high spin-dependent Seebeck coefficients of 150 µV/K [24].

Based on the spin dependent change of the thermopower of $\Delta V_{TP} = V_{TP}(AP) - V_{TP}(P)$ also the TMTP ratio of the devices can be calculated. When defining a TMTP ratio of the MTJ nanopillars as $TMTP = \Delta V_{TP}/ V_{TP}(P)$ one obtains values between 9 and 41 as listed in Table II. Note however that the TMTP ratio also contains a significant background contribution from all non-magnetic layers of the MTJ and the contacts. To remove these we have shortened the MgO barrier of some of the MTJ nanopillars by application of current stress. After barrier breakdown the TMR vanishes and the thermopower signal $V_{TP,short}$ becomes independent of field (cp. Ref. 11, Fig.2). $V_{TP,short}$ still scales linearly with $P_{heat}$ as shown in Fig. 2(f) (dashed line). Therefore it can be used to estimate the background thermopower of the remaining non-magnetic layers of the MTJ nanopillars. Subtracting this background from $V_{TP}(P)$ yields the TMTP contribution of the CoFeB/MgO/CoFeB MTJ. The resulting values of $TMTP_{MTJ} = \Delta V_{TP}/(V_{TP}(P) - V_{TP,short})$ of e.g. 72 % (MTJ-2) and 90 % (MTJ-3) are significantly higher than those listed in Table II.

Note that application of current stress did not result in a reliable barrier breakdown for all devices. While some of the devices still showed a TMR (and thus a transport contribution of the magnetic layers) others showed a significant resistance increase up to the MΩ range. Those devices could not be used for the above determination of $TMTP_{MTJ}$ and hence no values are given. In the future comparative thermopower measurements on reference stacks without MgO barrier should be used to more reliably determine the thermopower contributions of the non-magnetic layers in the MTJ stacks.

Concluding we have described TMTP measurements in CoFeB/MgO MTJ nanopillars. The temperature gradient across the MgO MTJ was determined using a calibrated heater line thermometer and heat flux modelling. Based on this large spin-dependent Seebeck coefficients were derived in agreement with ab-initio predictions. Considering non-mangetic background contributions allowed deriving the TMTP ratio of the 1.5 nm thick MgO barrier.

We acknowledge funding by the EU IMERA-Plus Grant No. 217257 and by the DFG Priority Program SpinCaT.

**TABLES:**

| Material | $\rho$ [$10^3$ kg/m$^3$] | $c_P$ [J/kg·K] | $\kappa$ [W/m·K] |
|---|---|---|---|
| Cu | 8.96 | 384 | 401 |
| Ru | 1.53 | 364 | 58.2 |
| Ta | 16.4 | 140 | 57.5 |
| Au | 19.3 | 129 | 317 |
| SiO$_2$ | 2.2 | 730 | 1.4 |
| | | | *bulk values from [17]* |
| MgO | 3.6 | 935 | 4* [19] |
| Co$_{60}$Fe$_{20}$B$_{20}$ | 8.2 | 440 | 87 |
| | | | *values from [10]* |
| Ta$_2$O$_5$ | 8.2 [22] | 322.9 [20] | 0.2* [23] |
| PtMn | 12.5 | 247 | 4.9 |
| | | | *values from [21]* |

*\* explicit thin film values which are different to the bulk values*

**Table I**

**N. Liebing et al.**

| N° | R$_P$ (Ω) | TMR (%) | TMTP (%) | ΔS (μV/K) |
|---|---|---|---|---|
| **1** | 322 | 79 | **32** | **250±40** |
| **2** | 243 | 88 | **17** | **120±20** |
| **3** | 195 | 110 | **32** | **270±40** |
| **4** | 389 | 134 | **41** | **390±60** |
| **5** | 397 | 137 | **30** | **210±30** |
| **6** | 213 | 116 | **37** | **330±50** |
| **7** | 216 | 88 | **9** | **60±10** |
| **8** | 207 | 95 | **31** | **290[i]±50** |

**Table II**

**N. Liebing et al.**

---

[i] It has been derived considering P$_{heat}$ = 60 mW. However, measurements were carried out at P$_{heat}$ = 51 mW

FIGURES :

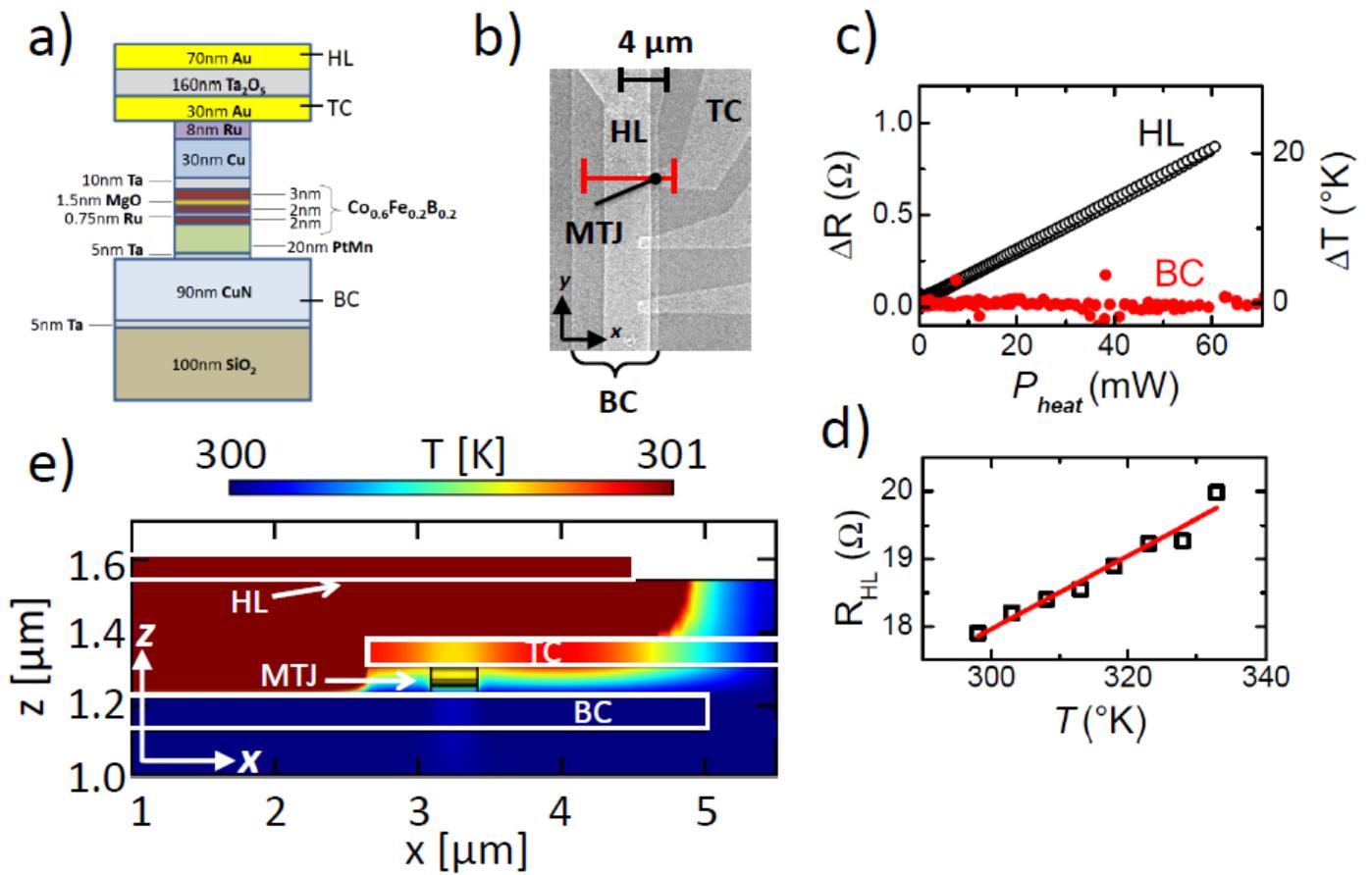

**Figure 1**

**N. Liebing et al.**

- 10 -

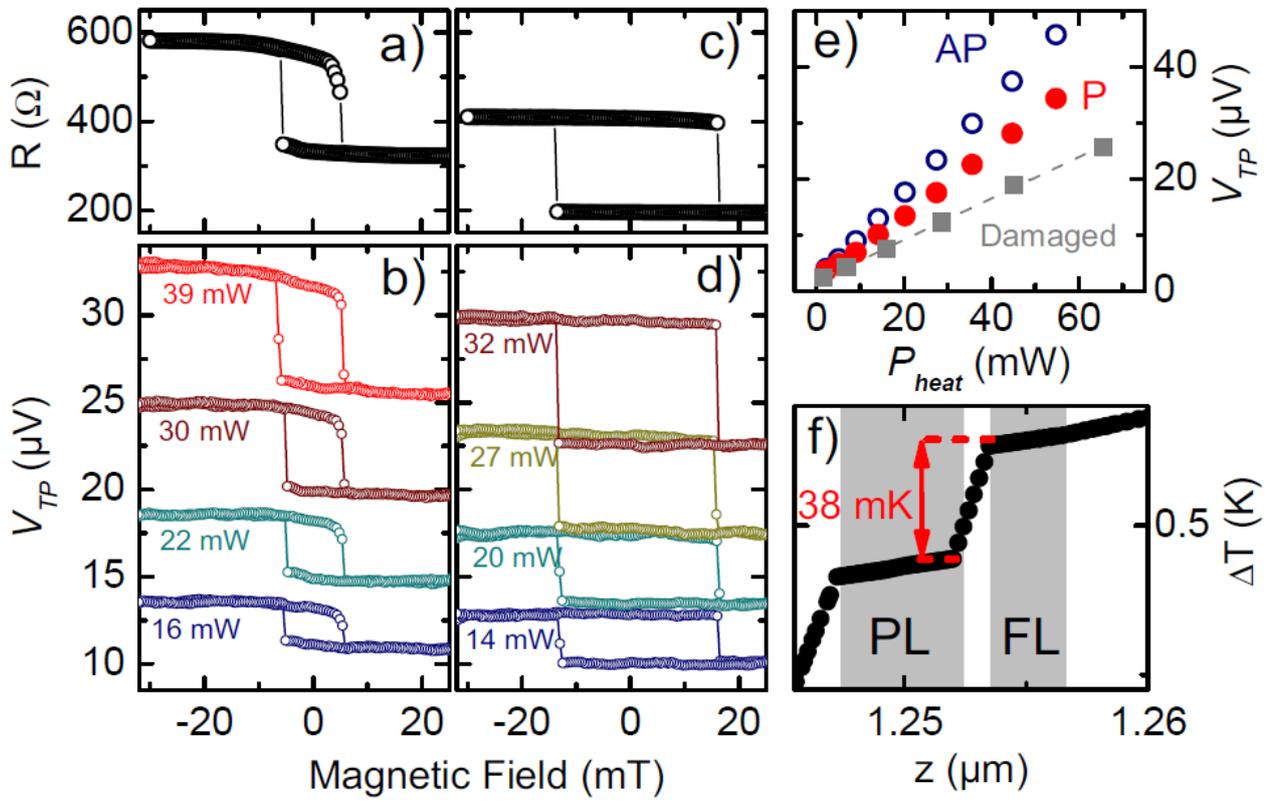

**Figure 2**

**N. Liebing et al.**

**FIGURE AND TABLE CAPTIONS:**

**Table I:** Material parameters with references used in heat flux simulations: density ρ, specific heat $c_P$, and heat conductance κ.

**Table II:** Resistance and thermoelectric properties of the measured MTJs: resistance in parallel state ($R_P$), TMR ratio, TMTP ratio and the spin-dependent Seebeck coefficient ($\Delta S$) with uncertainty resulting from HL temperature uncertainty.

**Fig. 1:** (a) Sketch of MTJ stack composition. HL, TC and BC, respectively refer to the heater line, electrical top contact of the MTJ and electrical bottom contact of the MTJ (b) SEM image of a typical device with the HL, BC, and TC. The position of the MTJ nanopillar is indicated. Red line indicates the cross section used for 2D simulations. (c) Resistance increase $\Delta R$ of heater line (open dots) and bottom contact (full dots) with heating power $P_{heat}$. Right scale: temperature increase $\Delta T$ as function of $P_{heat}$ of both HL and BC. (d) Measured temperature dependence of heater line resistance ($R_{HL}$). α is determined from linear fit. (e) Simulated temperature distribution in the 2D cross section. Position of HL, TC, MTJ, and BC are indicated.

**Fig. 2:** Easy axis TMR and TMTP loops of two typical devices MTJ-1 (a,b) MTJ-3 (c,d). (b) and (d) show TMTP loops at different DC heating powers $P_{heat}$. (e) $V_{TP}$ as function of $P_{heat}$ for parallel (P, red) and antiparallel (AP, black) orientation of MTJ-3. Gray dashed lines show the power heating dependence of the $V_{TP,\ short}$. (f) Simulated temperature profile across the MTJ nanopillar for $P_{heat}$ = 60 mW. $\Delta T_{MTJ}$ ~ 38±6 mK is derived.

# REFERENCES:

[1] I. Zutic, J. Fabian and S. Das Sarma, Rev. Mod. Phys. **76**, 323 (2004).

[2] J. Shi et al., Phys. Rev. B **54**, 15273 (1996).

[3] L. Gravier et al. Phys. Rev. B **73**, 024419 (2006). L. Gravier, et al. Ibid. B **73**, 052410 (2006).

[4] K. Uchida et al., Nature 455, 778 (2008).

[5] K. Uchida et al. Nat. Mater. **9**, 894 (2010).

[6] C. M. Jaworski et al. Nat. Mater. **9**, 898 (2010).

[7] A. Slachter et al. Nat. Phys. **6**, 879 (2010).

[8] G. E. W. Bauer, A. H. MacDonald, and S. Maekawa (Eds), Solid State Comm., Special Issue on Spin Caloritronics, Vol. 150, Issues 11-12, Pages 459-552 (2010).

[9] C. Chappert, A. Fert and F. Nguyen Van Dau, Nat. Mater. **6**, 813 (2007).

[10] M. Walter et al., Nature Mat. DOI: 10.1038/NMAT3076

[11] N. Liebing, S. Serrano-Guisan, K. Rott, G. Reiss, J. Langer, B. Ocker and H. W. Schumacher, Phys. Rev. Lett., in press and arXiv 1104.0537v2.

[12] J.-C. LeBreton et al. Nature **475**, 82 (2011).

[13] The stacks are sputter deposited in a Singulus NDT Timaris cluster tool on a Si wafer capped with 100 nm $SiO_2$.

[14] S. Serrano-Guisan et al. Phys. Rev. Lett., **101**, 087201 (2008).

[15] B Stahlmecke and G Dumpich, J. Phys.: Condens. Matter **19**, 046210 (2007).

[16] Comsol Multiphysics with Heat Transfer Model.

[17] William M. Haynes and David R. Lide. CRC Handbook of Chemistry and Physics. CRC Press, Boca Raton (Fla.), London, New York, 91th Edition 2010.

[18] A. J. Slifka, B. J. Filla, and J. M. Phelps, J. Res. Natl. Inst. Stand. Technol. **103**, 357 (1998).

[19] S.-M. Lee, David G. Cahill, and Thomas H. Allen. *Thermal conductivity of sputtered oxide films*. Phys. Rev. B **52,** 253 (1995).

[20] K.T. Jacob, Chander Shekhar, and Y. Waseda. *An update on the thermodynamics of $Ta_2O_5$*. The Journal of Chemical Thermodynamics **41,** 748, (2009).

[21] C Papusoi, R Sousa, J Herault, I L Prejbeanu, and B Dieny. *Probing fast heating in magnetic tunnel junction structures with exchange bias*. New Journal of Physics **10,** 103006 (2008).

[22] G. V. Samsonov. *The Oxide handbook*. IFI/Plenum, New York, 2d edition, 1982

[23] Z. L. Wu et al. . *Absorption and thermal conductivity of oxide thin films measured by photothermal displacement and reflectance methods*. Appl. Opt. **32**, 5660 (1993).

[24] Michael Czerner, Michael Bachmann, and Christian Heiliger, Phys. Rev. B **83**, 132405 (2011).